# Absorption bias: An ideal descriptor for radiation tolerance of nanocrystalline BCC metals

Liuming Wei[1,2], Zhe Zhao[1], Yonggang Li[1,2], Qirong Zheng[1,2], Chuanguo Zhang[1], Jingyu Li[1], Gaofeng Zhao[3], Bo Da[4] and Zhi Zeng[1,2]

[1] Key Laboratory of Materials Physics, Institute of Solid State Physics, HFIPS, Chinese Academy of Sciences, Hefei 230031, People's Republic of China

[2] University of Science and Technology of China, Hefei, 230026, People's Republic of China

[3] Institute for Computational Materials Science, School of Physics and Electronics, Henan University, Kaifeng, 475004, People's Republic of China

[4] Research and Services Division of Materials Data and Integrated System, National Institute for Materials Science, Namiki 1-1, Tsukuba, Ibaraki 305-0044, Japan

**Abstract**

To evaluate the radiation tolerance of nanocrystalline (NC) materials, the damage effects of Fe and W as typical body-centered cubic (BCC) metals under uniform irradiation are studied by a sequential multi-scale modelling framework. An ideal descriptor, the absorption bias (the ratio of the absorption abilities of grain boundaries (GBs) to interstitials (I) and vacancies (V)), is proposed to characterize the radiation tolerance of materials with different grain sizes. Low absorption bias promotes defects annihilation through enhancing I-V recombination and optimally tuning its competition with GB absorption. Thus, the lower absorption bias, the higher anti-irradiation performance of NC BCC metals is. Furthermore, by comprehensively considering the mechanical property, thermal stability and radiation resistance, nano-crystals are recommended for Fe-based structural materials but coarse crystals for W-based plasma-facing materials. This work reevaluates the radiation resistance of NC materials, resulting in new strategies for designing structural materials of nuclear devices through manipulating grain sizes.

**Introduction**

The design of high radiation resistant materials of fission/fusion devices is one of the key issues in the development of the nuclear industry (*1*). During long-time, high-energy neutron exposure, serious displacement damage and high concentrations of transmutation elements will be produced in nuclear materials (*2*). Point defects (like self-interstitial atoms (SIAs) and vacancies) will then agglomerate to form dislocation loops and large voids (*3*). The accumulation of these defects can alter the microstructures and result in adverse impacts such as the degradation of mechanical property, the decrease in thermal conductivity, and the increase of ductile-brittle transition temperature, eventually greatly constraining the durability of nuclear reactions. Therefore, compatible nuclear materials should be selected to combat these critical concerns. In current nuclear devices, body-centred cubic (BCC) metals and their alloys have been mostly adopted as the structural and plasma-facing components owing to their excellent mechanical and thermal properties as well as relatively high radiation/corrosion resistance (*4*–*7*). In the past decades, more attention has also been paid to the response of BCC metals under neutron irradiation (*8*–*10*). The performance of pure BCC metals as nuclear materials gradually approaches their limits. To meet the design requirements for future fusion reactors, material performance especially radiation resistance needs further improvement under intense irradiation. Thus, efforts are still necessary to understand the radiation tolerance of BCC metals, and to predict their long-term service performance in nuclear devices.

Recently, nano-crystallization is one of commonly potential ways to enhance the comprehensive performance of materials under irradiation (*11*–*13*). Both experimental and theoretical studies have shown that reducing grain size could significantly reduce defects in certain NC face-centred cubic (FCC) metals like copper (*13*, *14*), silver (*15*) and gold (*16*), and NC BCC metals like α-iron (Fe) (*17*), molybdenum (Mo) (*18*) and tungsten (W) (*19*–*21*). It is generally attributed to that the high density of GBs could act as efficient sites for absorbing intrinsic defects and emitting interstitials to annihilate vacancies near GBs (*13*). Nevertheless, some experimental studies have indicated that some materials such as pyrochlores $A_2Ti_2O_7$ (*22*), zirconia (*23*), SiC (*24*) and W (*25*) exhibit opposite anti-irradiation behaviours that the radiation tolerance of NC is relatively worse compared to coarse-grained (CG) materials under neutron or heavy ion irradiation. In fact, GBs could also absorb most of the fast mobile defects and hinder the recombination of SIAs and vacancies (I-V recombination), thereby reducing the radiation resistance of materials. Up to now, it still misunderstands the effect of grain size on the radiation tolerance of different kinds of materials under specific conditions. Furthermore, these experimental results can only show the macroscopic characteristics of materials, and lack a deep discussion of their microscopic mechanisms. On the other hand, in theory, various dynamic mechanisms and interaction rules of small size defects (typically less than several tens in number) with sinks like GBs in non-steady states at the time-scales less than nanoseconds (ns) have been simulated with atomistic methods including density functional theory (DFT), molecular statics and molecular dynamics (MD) (*26–28*). Whiles the steady state at infinite (long enough) time of systems with different grain sizes have been simulated with the steady state rate theory (RT) (*23*, *29*, *30*). In fact, the key dynamical processes from non-steady states at ns to the steady state at the infinite time are more critical for understanding the radiation tolerance of materials under continuous irradiation during their service process (*8*, *31*, *32*). Thus, to directly connect these two limit cases, the meso-scale methods such as kinetic Monte Carlo (*33–35*) and cluster dynamics (CD) (*36–38*) are usually used to simulate defect evolution and accumulation in irradiated BCC metals. In general, the multi-scale dynamic process of radiation damage needs to be systematically studied by using a multi-scale modelling framework. Moreover, many descriptors are used to characterize different macroscopic properties of materials such as the mechanical properties and thermal stability commonly described by the yield strength and Gibbs free energy, respectively. However, there is still no suitable descriptor to characterize the anti-irradiation abilities of materials till now, due to many complicated influencing factors (such as intrinsic properties of defects and structural features of materials). Therefore, it is necessary to seek for a simple and physically meaningful descriptor to evaluate the anti-irradiation performance from non-steady states to the steady state of metals with different grain sizes, and further to design high comprehensive performances of materials by adjusting grain size.

In this paper, we have proposed a descriptor, absorption bias, to describe the grain size-dependent anti-irradiation behaviours of BCC metals, respectively. A sequential multi-scale modelling framework coupling MD, CD and steady state RT models is adopted to study the radiation damage behaviours in typical BCC metals with different grain sizes. MD simulations show that the difference in migration energies determines the survival of intrinsic defects in Fe and W. CD simulations reveal that the effect of GBs on vacancy accumulation in BCC metals can be described by the absorption bias. Steady state RT simulations indicate that vacancy diffusivity plays a major role in the inherent radiation tolerance of base materials. Finally, the high-performance windows for Fe-based structural materials and W-based plasma-facing materials (PFMs) are proposed by comprehensively considering three descriptors of yield strength, Gibbs' free energy and absorption bias. This descriptor-based rational design strategy is targeted and enables us to efficiently design high-performance fusion structure materials.

**Absorption bias: a radiation tolerance descriptor**

In defect evolution, the migration energy $E_m$ determines the diffusivity of defects (like SIAs and vacancies) arising from the diffusion coefficients $D_{I/V} = D_{I/V}^0 \exp\left(-E_{I/V}^m / k_B T\right)$. Here we introduce the diffusion bias ($B_D$) as the ratio of the diffusion coefficients between SIAs (I) and vacancies (V),

$$B_\mathrm{D} = D_\mathrm{I}/D_\mathrm{V} = D_\mathrm{I}^0/D_\mathrm{V}^0 \exp(-\Delta E_\mathrm{m}/k_\mathrm{B}T). \tag{1}$$

This term determines the annihilation ability of defects through the formation of SIA clusters by quickly aggregating and reducing the recombination probability with vacancies. The diffusion bias is mainly affected by the intrinsic properties of materials ($\Delta E_\mathrm{m}$) and the temperature ($T$), where $\Delta E_\mathrm{m} = E_\mathrm{I}^\mathrm{m} - E_\mathrm{V}^\mathrm{m}$ scales the diversity of migration ability between SIAs and vacancies. Fig. 1A shows the temperature dependence of $B_\mathrm{D}$ for seven typical BCC metals (Fe, V, Nb, Cr, Ta, Mo and W) which are commonly used for structural materials in nuclear devices (*39*). It can be found that the values of the diffusion bias decay exponentially with increasing temperature. The decay rates of diffusion bias with temperature for Ta, Mo and W are faster than those of Fe, V, Nb and Cr, due to their larger $\Delta E_\mathrm{m}$. In contrast, for Fe, V, Nb and Cr with smaller $\Delta E_\mathrm{m}$, $B_\mathrm{D}$ gradually approaches a constant with increasing temperature.

In practice, the absorption rate of defects at GBs in polycrystalline (PC) materials is directly related to the diffusion coefficients of defects and the sink strengths of GBs. Thus, we introduce the absorption bias ($B_A$) to characterize the ratio of the absorption abilities of GBs to SIAs and vacancies as

$$B_\mathrm{A} = \left(K_\mathrm{GB}^\mathrm{I}/K_\mathrm{GB}^\mathrm{V}\right)B_\mathrm{D}, \tag{2}$$

where $k_\mathrm{sc}^\mathrm{I}$ and $k_\mathrm{sc}^\mathrm{V}$ are the sum of the sink strengths of all the intragranular sinks of SIAs and vacancies but GBs. Among the several existed models of GB sink strength to mobile defects (*40–42*), it is reasonable to choose the cellular model to describe the absorption effect of uniform GBs on mobile defects, because the cellular sink strength is the preferable one for regular distributed GBs, as given by (*40*),

$$K_\mathrm{GB}^\theta = k_\mathrm{sc}^\theta \left(\frac{\sqrt{k_\mathrm{sc}^\theta}d}{2}\coth\frac{\sqrt{k_\mathrm{sc}^\theta}d}{2}-1\right)\left(1+\frac{k_\mathrm{sc}^\theta d^2}{12}-\frac{\sqrt{k_\mathrm{sc}^\theta}d}{2}\coth\frac{\sqrt{k_\mathrm{sc}^\theta}d}{2}\right)^{-1}, \tag{3}$$

which is related to the grain size ($d$) and the sum of the sink strengths of all the intragranular sinks but GBs ($k_\mathrm{sc}^\theta$).

There are two limiting cases for $B_A$ when defects loss to intragranular sinks are very small compared to that to GBs ($\sqrt{k_\mathrm{sc}^\theta}d \to 0$) and dominate over that to GBs ($\sqrt{k_\mathrm{sc}^\theta}d \to \infty$), Eq. (3) tends to $K_\mathrm{GB}^\theta = 60/d^2$ and $K_\mathrm{GB}^\theta = 6\sqrt{k_\mathrm{sc}^\theta}/d$, and thus Eq. (2) becomes

$$B_A = B_D, \qquad (\sqrt{k_{sc}^\theta} d \to 0) \qquad (4)$$

and

$$B_A = \sqrt{k_{sc}^I / k_{sc}^V} \cdot B_D, \qquad (\sqrt{k_{sc}^\theta} d \to \infty). \qquad (5)$$

The grain size $d$ dependence $B_A$ by Eqs. (2), (4) and (5) is plotted in Fig. 1B. Here, the values of $k_{sc}^{I/V}$ could be extracted from CD calculation under uniform damage in W at 400 K, which will be presented later. Apart from the two limiting values (Eqs. (4) and (5)) shown as upper dash line and lower dash dot line in Fig. 1B, $B_A$ decreases rapidly as $d$ increases. We will show later that the diffusion bias and adsorption bias could be used as good descriptors to well characterize the radiation tolerance of base materials and materials with different grain sizes, respectively.

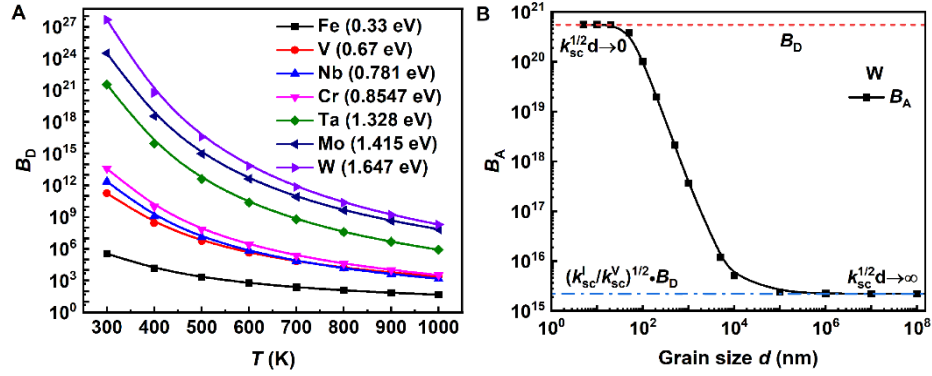

**Fig. 1. Diffusion bias and absorption bias.** (**A**) Temperature dependence of $B_D$ for seven BCC metals whose $\Delta E_m$ are given in the brackets. (**B**) Grain size dependence of $B_A$ of W at 400 K. The red dash and blue dash dot lines are the two limiting values of $B_A$ (Eqs. (4) and (5)).

**Results**

The radiation damage process of nuclear materials includes three states: the initial irradiation in the non-steady state at about ns, non-steady states of the dynamic behaviours of defects under a long-time irradiation, and the final steady state of materials after irradiation. In the following, the multi-scale modelling framework coupling MD, CD and steady state RT models is adopted to study defect behaviours in BCC metals during this process. In general, the performance reduction under irradiation is proportional to the densities and sizes of defects (*19*). In particular, vacancies and their clusters play an important role in the mechanical properties (swelling, hardening and embrittlement), thermal properties, hydrogen (H)/helium (He) retention of materials. Therefore, we choose the concentration of vacancies as a criterion for considering the radiation tolerance of materials here.

**Defect behaviours near GBs in non-steady states at nanoseconds**

In order to show defect behaviours at the initial stage of irradiation, we used the MD method (as described in the Method section) to simulate the annihilation of point defects by GBs in Fe and W at ns. The detailed settings of MD calculations can be seen in Section S1. The space-distributions of SIAs and vacancies near typical ∑5 tilt GB and their numbers at different distances apart from the GB after 2 ns evolution in Fe and W at different temperatures are shown in Fig. 2. Obviously, more SIAs than vacancies are accumulated in GBs for both Fe and W. While more vacancies retain in the grain interior of W than that of Fe. It is mainly because the diffusivity of SIAs (with the migration energies of 0.34 eV for Fe and 0.013 eV for W) is typically higher than that of vacancies (with the migration energies of 0.67 eV for Fe and 1.66 eV for W). In other words, there is a high diffusion bias ($B_D$) especially in W. In addition, with lower $B_D$ and the corresponding lower absorption of SIAs in GBs, the I-V recombination coefficient in Fe (0.774) is higher than that in W (0.642). As shown in Fig. 2A, for Fe, most vacancies are annihilated by GBs and a defect-free zone is thus formed within a distance of about 1 nm from the GB after 2 ns at 600 K. Thus, the anti-irradiation performance of NC Fe is improved by absorbing most of SIAs and vacancies in GBs. On the contrary, for W (Fig. 2B), SIAs quickly diffuse and are absorbed in GBs, while vacancies with very low diffusivity even at a high temperature of 873 K can only recombine with little SIAs in grain interior instead of annihilation in GBs. Additionally, there is a high density of GBs in NC W, and therefore more SIAs will be absorbed by GBs but less recombination with vacancies in the grain interior, which will lead to a higher residual vacancy density in the grain interior of NC W than that of CG W. Consequently, the anti-irradiation ability of NC W becomes even worse by increasing the density of GBs at ns time-scale. $B_D$ could well describe the radiation tolerance of different base materials, which is only determined by the diffusivities of the intrinsic defects.

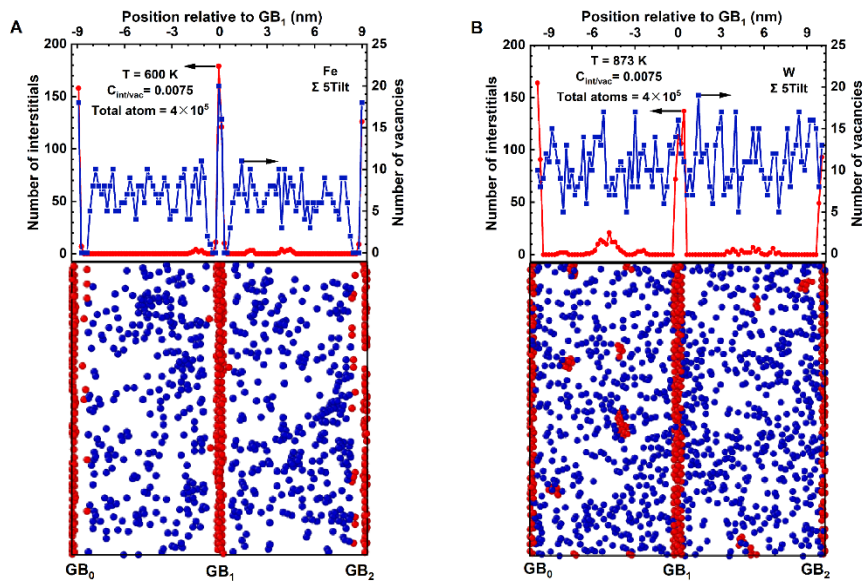

**Fig. 2. Defects distribution by MD simulation.** Space distributions (bottom) and number (top) of SIAs and vacancies near ∑5 tilt GB projected from the abscissa axis after 2 ns evolution in Fe (**A**) and W (**B**). The defects were visualized using the OVITO tool (*43*). SIAs and vacancies are represented by red and blue spheres, respectively.

While the systems studied here are far from the steady state at the very short time-scale of ns. In practice, systems are still in non-steady states because defect production is much faster than defect annihilation under continuous irradiation. The long-term dynamic evolution of defects from non-steady states at ns to the steady state should be even more critical in understanding the radiation tolerance of practical materials with intrinsic sinks like GBs. Further exploration is still needed to describe the anti-irradiation abilities of base materials by $B_D$ in non-steady states under typical conditions. Moreover, because $B_D$ does not take into account the effect of GBs, it is necessary to use $B_A$, which contains the absorption strength of GBs to defects to study the anti-irradiation behaviours of BCC metals with different grain sizes.

**Defect evolution in non-steady states**

We adopt the CD model (as described in the Method section) (*44, 45*) to systematically simulate the effect of diffusion bias and absorption bias on the radiation tolerance of base materials and materials with grain sizes from nm to infinite at the time-scale from ns to s. The verification of our CD model can be seen in fig. S2. In the following, the effect of grain size on total vacancy accumulation in Fe (lowest $B_D$) and W (largest $B_D$) in non-steady states is studied within this approach. The simulation conditions are set as the production rate of Frenkel pairs of $10^{-6}$ displacements per atom/s (dpa/s) with duration time up to 100 s. The grain size and temperature dependence of relative vacancy concentration ($C_V/C_V^0$, the vacancy concentration $C_V$ of PC materials normalized to $C_V^0$ of the single crystal (SC)) in Fe and W are given in Fig. 3, which has demonstrated a radiation resistant window related to grain size and temperature. The black solid line represents the total vacancy concentration in the SC, which divides the whole region into the radiation resistant ($C_V/C_V^0 < 1$) and non-radiation resistant regions ($C_V/C_V^0 > 1$). It can be found that the radiation resistant window of Fe is much larger than that of W. As expected, the radiation tolerance of NC Fe is improved when the temperature is above 450 K. While the radiation resistant window of W is very small, which shows no advantage on vacancy annihilation even at the high temperature of 900 K (the typical operating temperature of W-based PFMs). The variations of vacancy concentration of Fe and W with grain size at typical temperatures (400 and 900 K) have been shown in fig. S3. There is a typical heavy damage region at low temperatures and at nano-scale (dark red region in W), mainly caused by the extremely serious effects of high $B_D$ for W. For example, the damage in W with the grain size smaller than 100 nm and under the temperature lower than 700 K is two orders heavier than that in SC W. In addition, the heavy damage region for W is larger than that for Fe, which is mainly caused by the larger $B_D$ of W. Thus, $B_D$ can be used to evaluate the intrinsic anti-irradiation abilities of base metals. A small difference in the migration energies of SIAs and vacancies for Fe leads to a small $B_D$ and finally to a slight effect of $B_A$ on the segregation of SIAs and vacancies when increasing temperature or decreasing grain size, resulting in a wide radiation resistant window. In contrast, a large difference in migration energies of SIAs and vacancies for W leads to a large $B_D$ and finally to a significant effect of $B_A$ on the segregation of SIAs and vacancies when temperatures are not high enough or decreasing grain size, resulting in a narrow radiation resistant window. Increasing $B_A$ could decrease the ratio between I-V recombination in grain interior and the defect annihilation at GBs dramatically when decreasing temperature or grain size, causing a smaller radiation resistant window for W than for Fe. In general, for practical materials with different grain sizes, the effects of vacancy diffusivity ($D_V$) and $B_A$ are related to the temperature ($T$) and grain size ($d$), respectively. High temperatures can enhance $D_V$, which will increase the probabilities of I-V recombination and defect diffusion to GBs. For different service temperatures in special service environments, by modifying $d$ of materials to tune $B_A$, one could further control or improve the radiation tolerance of materials. Therefore, the absorption bias can be chosen as an *ideal descriptor* of the anti-irradiation abilities of BCC metals with different grain sizes.

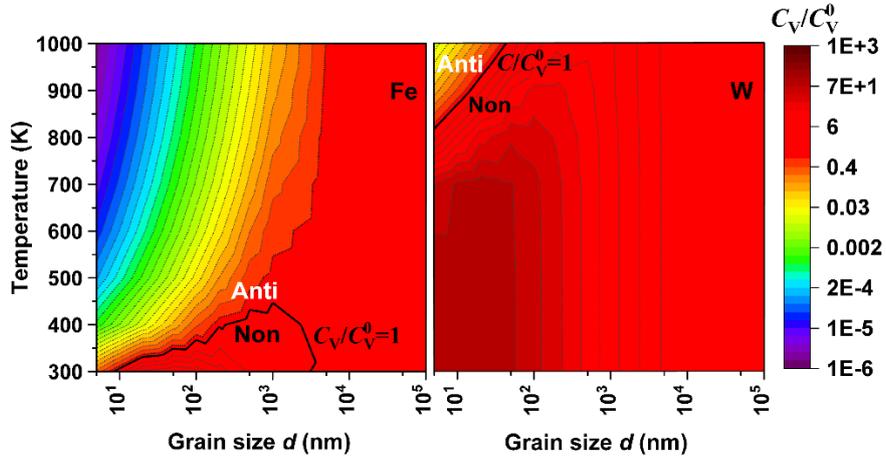

**Fig. 3. Radiation resistant windows predicted by the CD model.** Grain size and temperature dependence of normalized vacancy concentration in Fe and W under uniform irradiation of 100 s with the production rate of Frenkel pairs of $10^{-6}$ dpa/s. The $C_V$ of PC material normalized to that of the SC $C_V^0$ in Fe and W is used to divide the radiation resistant window ($C_V/C_V^0 =1$ (the black line), that is, $C_V/C_V^0 <1$ is the radiation resistant region (Anti), $C_V/C_V^0 >1$ is the non-radiation resistant (Non)).

In order to correspond to experimental results of the defect distribution and thermodynamic properties of materials after irradiation, we further added 1000 s annealing after irradiation for the same case of W in Fig. 3. Defects would diffuse to and be absorbed by GBs during annealing. In the low temperatures (vacancies are not activated), the number of SIAs absorbed by GBs decreases and more SIAs participate in the I-V recombination with increasing grain size, which promotes the annihilation of vacancies. In the high temperature, vacancies begin to be activated and absorbed by GBs with decreasing grain size (fig. S4). Thus, the vacancy accumulation is reduced and the radiation resistant window of W is enlarged as shown in Fig. 4. This theoretical predicted radiation resistant window could be verified by comparing the theoretical simulation results with the experimental anti-irradiation abilities of pure W and its alloys with different grain sizes under heavy ion/neutron irradiation (*19–21*, *25*, *46*, *47*). Five experiments support our theoretical predictions on the anti-irradiation ability of NC W compared to CG W. In these experiments, researchers indicated that NC W has better (*19–21*, *47*) or worse (*19*, *25*) anti-irradiation ability than CG W at high (*19–21*, *47*) or low (*19*, *25*) temperatures, respectively, by comparing their mechanical properties (*19*, *20*) and defect densities (*19*, *21*, *25*, *47*) after ion (*19*, *20*, *25*, *47*)/neutron (*21*) irradiation. Only one experiment (*46*) presented an adverse conclusion that the anti-irradiation ability of ultra-fine grained (UFG)-W-TiC is better than pure W at 563 K. However, UFG-W-TiC presented in this experiment is not a pure W, where TiC dispersed microstructure could contribute more resistance to radiation hardening (*46*). In addition, more evidences can be found that the anti-irradiation ability (lower defect density and hardness changing) of NC Fe is better than CG Fe (*48*, *49*), confirming the universal applicability of our predictions in other metals. Furthermore, some simulation results also indicated the same conclusion with us that the concentration of vacancies decreases with decreasing grain size when vacancies are activated in W (*34*, *47*). In general, the theoretical prediction should also be universal for other BCC metals.

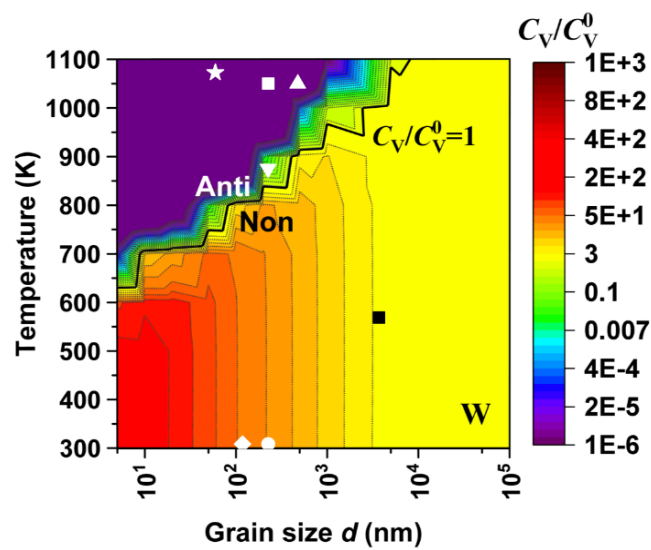

**Fig. 4. Radiation resistant window after annealing.** Verification of radiation resistant window of CD predictions with experiments (*19–21*, *25*, *46*, *47*) for W under uniform

irradiation of 100 s with the production rate of Frenkel pairs of $10^{-6}$ dpa/s and 1000 s annealing. The hollow and solid symbols indicate the experiments supporting (□○△▽◇☆) (*19–21, 25, 47*) and contradicting (■) (*46*) our predictions, respectively.

**Anti-irradiation behaviour in the steady state**

In fact, systems are always in non-steady states under practical conditions, because a long enough time is needed to reach the steady state especially for metals with high defect migration energies at low temperatures (like that about $10^{27}$ s of W with the grain size of 10 μm even at 300 K as shown in fig. S5). In the ideal case, after long enough time annealing, all kinds of defects could be fully absorbed by GBs, whether the $B_D$ and $B_A$ still affect defect accumulation in the steady state? The steady state RT is introduced here (as described in the Method section) to discuss the steady state for a long enough time of defect accumulation in typical BCC metals with different grain sizes. For RT equations in the steady state ($\partial C_I/\partial t = \partial C_V/\partial t = 0$, where $C_I$ and $C_V$ are the concentration of SIAs and vacancies, respectively), the formula of vacancy concentration can be solved analytically as (*23, 30*),

$$C_V = \frac{1}{B_V}\left(\sqrt{A_V^2 d^{-4} + 2B_V G} - A_V d^{-2}\right), \tag{6}$$

where $A_V = 57.6 D_V$, $B_V = (8\pi/a^2) D_V (D_V/D_I)(1 + D_I/D_V)$, and $G$ is the production rate of Frenkel pairs.

There are two asymptotes for the change of $C_V$ with grain size $d$ in the limits of small grain size and SC that correspond to the contributions of fully GB absorption and I-V recombination, respectively. The intersection of these two asymptotes is defined as the critical grain size $d_c$ (as shown in the insert fig. S6) (*30*),

$$d_c = \sqrt{28.8a}/\pi^{1/4} (D_V/G)^{1/4}, \tag{7}$$

to characterize the anti-irradiation abilities of metals under different service conditions. $d_c$ represents the upper grain size limit at which vacancy accumulation starts to be lower than that of SC, which shows a 1/4 power law of the anti-irradiation abilities of materials in the steady state. Materials with grain size less than $d_c$ have the lower values of $C_V/C_V^0$ and thus a better anti-irradiation ability (fig. S6).

In the steady state, $B_A$ is equal to $B_D$ due to the equivalent value of $K_{GB}$ for SIAs and vacancies. The critical grain size $d_c$ and the radiation resistance of materials are no longer affected by $B_D$ but only by $D_V$ (where $D_V$ is related to $E_V^m$ and $T$) and $G$. Therefore, by selecting base metals with appropriate $D_V$ (or $E_V^m$), new materials with high radiation resistance could be designed by manipulating $d$ to endure different service environments (like $T$ and $G$). A high radiation resistant material should have low $E_V^m$ and small $d$ in the steady state. Meanwhile, the lower $G$ and higher $T$, the better anti-irradiation ability of the material. Furthermore, we take Fe and W as examples to study the effect of grain size on radiation damage under different temperatures in the steady state (for details see Section S7), which shows that the high anti-irradiation abilities of Fe and W appear in the conditions of nano-scales and at high temperatures.

**Discussion**

**Key factors in the descriptor of absorption bias**

From the analysis above, it is clear that the deviation in anti-irradiation ability between Fe and W with different grain sizes in non-steady states originates from the difference in their absorption bias $B_A$. Furthermore, the absorption bias as a descriptor is mainly related to the GB sink strength and defect diffusivity of the material. In addition, the diffusivity of a defect is related to its migration energy and the temperature. Therefore, the absorption bias dominates the anti-irradiation ability of materials with different grain sizes essentially. We now analyze the variations of $B_A$ *v.s.* grain size $d$ by comparing the behaviours in different BCC metals at a specific temperature and a specific metal at different temperatures. In the case of different BCC metals under uniform irradiation at 400 K, the change of $B_A$ as a function of $d$ is shown in Fig. 5A. For metals with high vacancy migration energies (like Ta, Mo and W) (*39*), with increasing $d$, $B_A$ gradually decreases from a constant value (a plateau value in a small grain size range) after $d$ reaching the inflection point $d_1$, and then decreases to another constant value (a plateau value in a large grain size range) after $d$ reaches the other inflection point $d_2$. At 400 K, SIAs are highly mobile, while vacancies are nearly immobile. When $d < d_1$, $B_A$ is equal to $B_D$ because both SIAs and vacancies can fully interact with GBs to reach their respective stable states of maximum absorption ($k_{sc}d \to 0$). For this case, the larger $B_D$, the higher plateau value is. Moreover, the higher vacancy migration energies ($E_V^m(\text{W}) < E_V^m(\text{Mo}) < E_V^m(\text{Ta})$), the smaller the critical grain size $d_1$ is ($d_1(\text{W}) < d_1(\text{Mo}) < d_1(\text{Ta})$). When $d_1 < d < d_2$, the decrease of $B_A = \left(K_{GB}^I\left(k_{sc}^I, d\right) \big/ K_{GB}^V\left(k_{sc}^V, d\right)\right) B_D$ with increasing $d$ is the synergistic effect of grain size ($d$), internal sink strength ($k_{sc}$) and diffusion bias ($B_D$). When $d > d_2$, the absorption ability of GBs on mobile defects nearly vanishes so that $B_A$ approaches the case of the single crystal. At this moment, according to Eq. (5), $B_A$ no longer changes with $d$ and tends to the constant limit ($\sqrt{k_{sc}^I/k_{sc}^V} \cdot B_D$) due to the constant value of $k_{sc}^I/k_{sc}^V$ (for details see Section S8). For metals with low vacancy migration energies (like Fe, V, Nb and Cr) (*39*), $B_A$ does not change with $d$.

When $k_{sc}d \to 0$ caused by $d \to 0$, $B_A$ is equal to $B_D$ due to both SIAs and vacancies can fully interact with GBs, which is similarly to metals with high vacancy migration energies. With increasing $d$, on the one hand, the effect of $d$ itself on the change of GB absorption abilities to SIAs and vacancies is nearly the same. On the other hand, the change of $k_{sc}^I$ and $k_{sc}^V$ is almost the same. Therefore, the synergistic effects contribute to that the ratio of $K_{GB}^I(k_{sc}^I,d)/K_{GB}^V(k_{sc}^V,d)$ no longer varies with $d$. When $d \to \infty$, $B_A$ ($\sqrt{k_{sc}^I/k_{sc}^V} \cdot B_D$) tends to $B_D$ since the sink strength of single defects (both SIAs and vacancies) are nearly dominated by DLs ($k_{sc}^I/k_{sc}^V \to 1$). Therefore, $B_A$ is equal to $B_D$ within all grain size ranges. Overall, different base materials have different vacancy diffusivities even at the same temperature, and thus different diffusion bias. Two key factors, vacancy migration energy $E_V^m$ as an intrinsic defect property and grain size $d$ as a structure feature, predominantly influence $B_D$ and/or $k_{sc}^I/k_{sc}^V$, and thus $B_A$, which determine the radiation tolerance of different materials.

In the following, we take W as an example to explore the effect of temperature ($T$, one of the irradiation conditions) on $B_A$, as shown in Fig. 5B. The temperature affects the variation of $B_A$ with grain size is similar to that $E_V^m$ affects $B_A$ also by changing vacancy diffusivities with different metals at the same temperature. Except for no plateau value of $B_A=B_D$ when $k_{sc}d \to 0$ at 300 K. It is because at 300 K, the limit state that defects start to fully interact with GBs hardly be reached for the low vacancy diffusivity. With increasing temperature, vacancies are gradually activated, which increases the probability of vacancy interaction with GBs and reduces $B_A$. Thus, temperature is another key factor to affect the absorption bias and thus the radiation resistance of materials by changing the diffusivity of vacancies.

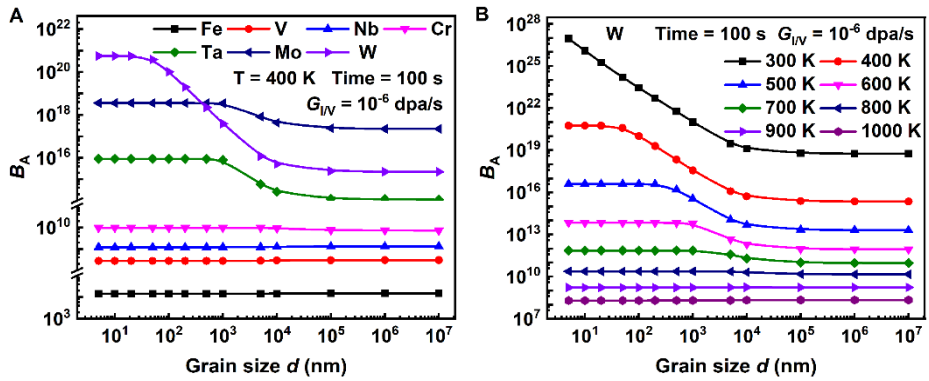

**Fig. 5. Effects of migration energy and temperature on absorption bias.** (**A**) $B_A$ as a function of $d$ in BCC metals under uniform irradiation of 100 s with the production rate of Frenkel pairs of $10^{-6}$ dpa/s at 400 K. (**B**) $B_A$ as a function of $d$ in W at different temperatures under uniform irradiation of 100 s with the production rate of Frenkel pairs of $10^{-6}$ dpa/s.

Upon the above analysis, the anti-irradiation abilities of materials could be derived by inspecting the statistical critical grain size ($d_c'$) when vacancy concentration reaches the value of SC in the steady state and the inflection point ($d_1$) at which defects start to fully interact with GBs in non-steady states. Fig. 6 shows that both $d_c'$ in the steady state and $d_1$ in non-steady states increase linearly with $(D_V/G)^{1/4}$, which satisfies the 1/4 power law relationship as in Eq. (7). Furthermore, $d_c'$ is approximately equal to $d_1$ for different materials at different temperatures. When $d$ is smaller than $d_c'$ or $d_1$, SIAs and vacancies can fully interact with GBs and reach the plateau value in different materials at the same temperature or the same material at different temperatures. The larger $d_c'$ or $d_1$, the better radiation resistance of the material is. No matter in the steady state or non-steady states, there should be a unified critical grain size ($d_c' \approx d_1$) that determines the anti-irradiation abilities of materials. Therefore, we could evaluate the radiation tolerance of materials in practice just by referring to the radiation tolerance of materials in the steady state.

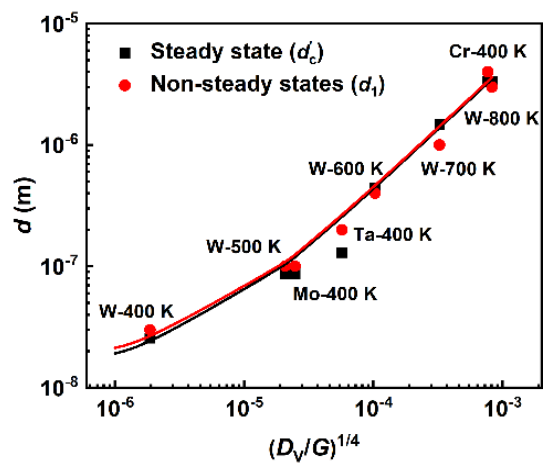

**Fig. 6. Critical grain sizes.** The comparison of the relationship of $d_c'$ and $d_1$ with $(D_V/G)^{1/4}$ in both the steady state and non-steady states.

In conclusion, the radiation resistance of metals with different grain sizes is directly related to the intrinsic properties (like $E_V^m$) and microstructures (like $d$) of materials, as well as irradiation conditions (like $T$). As an ideal descriptor, $B_A$ can reflect the effects of all these key factors. The lower $B_A$, the better the radiation resistance of a material is. Therefore, in the material design, appropriate materials should be intensively selected with low $B_D$ when only the anti-irradiation abilities of different base materials are compared and the material structures like $d$ should also be adjusted with low $B_A$ to improve their properties with superior radiation resistance under a specific service environment. Researchers have also found that the anti-irradiation ability of materials can indeed be improved when reducing the diffusion bias of intrinsic defects through doping impurity sinks to make $D_I^{eff} \approx D_V^{eff}$ (where $D_{I/V}^{eff}$ are the effective diffusivities of SIAs and vacancies) (*50*).

**Anti-irradiation mechanisms of NC versus CG materials**

A new mechanism of radiation resistance of NC versus CG materials could be derived here. In fact, $B_A$ can tune the optimal competition between two important self-healing processes of I-V recombination and GB absorption. At high $B_A$, the probability of I-V recombination in grain interior will be reduced due to more SIAs diffusing to and absorbed by GBs and more vacancies left in grain interior in NC than that in CG. It thus implies that the radiation tolerance in NC is typically worse than that in CG. In contrast, at low $B_A$, the anti-irradiation ability of material can be effectively improved due to the enhanced probability of I-V recombination in grain interior and more mobile defects can be absorbed by a high density of GBs in NC than that in CG. In this case, the radiation tolerance of NC is typically better than that of CG materials. In addition, the vacancy diffusivity is determined by material properties ($E_v^m$) and service temperatures (*T*), which tunes both the efficiency of GB annihilation and I-V recombination in grain interior. Thus, higher vacancy diffusivity can induce better self-healing ability. To sum up, base materials with excellent anti-irradiation ability should possess two critical criteria, that is, low migration energies of both SIAs and vacancies as well as a small difference between them (low $B_D$). These two criteria dominate the enhancement of GB self-healing through increasing defect diffusivity and the fraction of I-V recombination self-healing through decreasing $B_D$, respectively. In addition, decreasing grain size would enhance GB self-healing but meanwhile suppress I-V recombination self-healing. For a certain material, it is necessary to adjust grain size to seek the maximum of total self-healing ability. Except for the diffusion coefficients, the absorption rates of GBs also depend on the sink strength and the dynamic concentration of mobile defects. $B_D$ is an intrinsic defect property of materials, while $B_A$ is an extrinsic property due to sinks like GBs. Thus, in view of intrinsic properties (like $B_D$) of materials, the radiation tolerance of Fe is much better than that of W under the same irradiation condition. Considering the effect of grain size and $B_A$ on radiation resistance, NC Fe and CG W should be selected in practice. The diffusion bias and absorption bias could also describe the anti-irradiation performance of other base materials and materials with different grain sizes.

In the design of materials for fusion devices, it should simultaneously consider not only the radiation resistance but also the mechanical property and thermal stability of materials. These properties of materials with different grain sizes (*d*) under the service temperatures (400 K for Fe and 700 K for W) follow several empirical theories, as given in Fig. 7. Here the yield strength (mechanical property) is obtained by the Hall-Patch relationship (*51*), $\sigma = \sigma_0 + kd^{-1/2}$, where $\sigma_0$ is the friction stress (43 and 640.33 MPa for Fe and W, respectively) (*52*), and *k* is the strengthening coefficient (0.71 and 0.79 MN/m$^{3/2}$, respectively) (*52*). The thermal stability is described by the Gibbs free energy relationship (*23*), $G_{NC} = G_0 + 3V_m\gamma/d$, where $G_0$ is the Gibbs free energy of pure metals (796722.7 and 1276152.9508 J/mol for Fe and W calculated by us using the DFT software of Phonopy (*53*), respectively), $V_m$ is the molar volume and $\gamma$ is the specific Gibbs free energy of GBs (0.428 and 1.02 J/m² for Fe and W, respectively) (*54*). In this study, the anti-irradiation ability is described by $B_A$ simulated by the CD model under typical irradiation conditions. As shown in Fig. 7, both for Fe and W, the thermal stability (red line)/yield strength (black line) increases/decreases with increasing grain size *d* and approaches to the SC state finally. However, the variation of anti-irradiation ability *v.s*. *d* of W is quite different from that of Fe, seeing the blue dashed lines. Where the anti-irradiation ability of Fe slightly decreases with increasing grain size, and the anti-irradiation ability of W in terms of $B_A$ only begins to work at a critical grain size $d_1$, then increases until the performance of SC W as increasing grain size *d*. In considering the specific situation in fusion devices, in addition to the common requirement of anti-irradiation performance, Fe-based structural materials are required to yield strengthening but W-base armor materials are required for higher thermal stability. Furthermore, the comprehensive performance (purple short dash lines in Fig. 7) can be estimated by setting, for example, the ratio of mechanical properties to thermal properties to anti-irradiation properties as 4:2:4 and 2:4:4 for Fe and W, respectively. The high-performance window (the grey-blue range) for Fe- and W-based materials are then obtained as shown in Fig. 7. For Fe-based structural materials, its high-performance window is in the grain size range of tens nm, which is consistent with the experimental and theoretical results that NC Fe has well performance (*17*, *55*). For W-based PFMs, its high-performance window is in the grain size range of tens to hundreds μm, which is consistent with the grain size of W adopted in ITER or commercial W (*56–58*). Moreover, when considering their relative superiority in both comprehensive performances mentioned above and H/He retention (*59*), CG W should also be selected as the candidates of PFMs for fusion devices in practice.

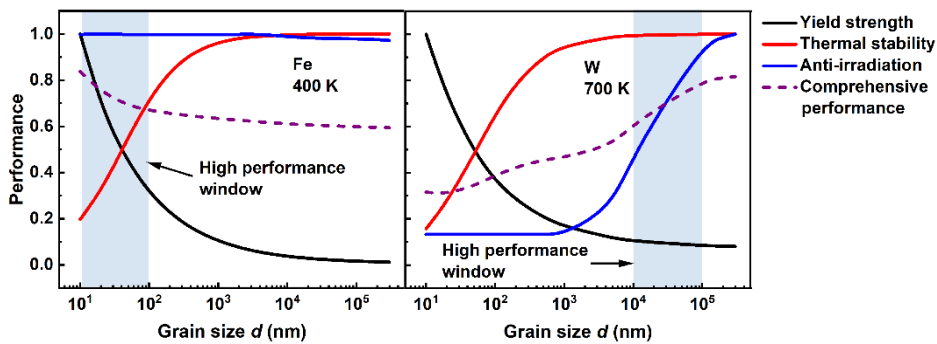

**Fig. 7. Comprehensive performance prediction for Fe- and W-based materials.** High-performance windows for Fe and W, when considering the synergy of thermal stability, mechanical property (yield strength), anti-irradiation ability with different grain sizes. Here, the performance is normalized to the best value.

In summary, we used a multi-scale modelling framework to systematically study the radiation tolerance of materials. We proposed the descriptors of diffusion bias $B_D$ to describe the radiation tolerance of different base materials and the absorption bias $B_A$ to describe the radiation tolerance of materials with different grain sizes. Base materials with low diffusion bias and NC metals with low absorption bias have good radiation resistance. The absorption bias is an ideal descriptor that can incorporate essential properties on both intrinsic factors (defect diffusivity) of materials and adjustable factors (grain size) in design, which provides a criterion for determining whether the nano-crystallization could improve the anti-irradiation abilities of materials. Thus, we recommend that nano-crystals should be selected for structure materials because of their better anti-irradiation abilities (low absorption bias) of metals like Fe, V, Nb and Cr. In contrast, coarse crystals should be selected for such as PFMs because of the high anti-irradiation abilities (low absorption bias) and low H/He retention rates of metals like Ta, Mo and W. Taking the properties of thermal stability, mechanical property and anti-irradiation ability into account, NC and CG materials are the best candidates for designing structure materials and PFMs, respectively. Our findings provide new insights into the micro-mechanisms of materials with different grain sizes under irradiation, which will be much helpful for the design and selection of materials under extreme conditions.

**Methods**

**Multi-scale modelling framework**

The dynamic evolution of defects in materials is a multi-scale and multi-micro mechanism phenomenon, which includes defect production, diffusion, reaction and accumulation. A sequential multi-scale modelling framework has been established to describe this process by coupling the atomistic (DFT and MD), Monte Carlo (IM3D (*60*)), CD (IRadMat (*44*, *61*)) and steady state rate theory (RT) models (*23*). In this work, this multi-scale modelling framework (MD+CD+RT) was adopted to study the long-term dynamic evolution behaviours of intrinsic defects under uniform irradiation.

**Molecular dynamics simulation**

At the beginning of irradiation from picoseconds to nanoseconds, the distribution behaviours of point defects near GBs in non-steady states at ns can be simulated by the open-source MD code LAMMPS (*62*). The Finnis-Sinclair type Fe potential (*63*) and Derlet-Nguyen-Manh-Dudarev type W potential (*64*) with the modification of the electron density function[31] were implemented in this study. These two potentials can reproduce correct point defect structures and experimental defect threshold energies (*63*, *65*). The initial configurations of energy minimization structures were relaxed within the NVT ensemble (constant number of atoms, volume, and temperature) for 2 ns with a time step of 1 fs. To reduce statistical error, 10 independent simulations of defect annealing were performed using a random distribution of $3\times10^3$ initial point defects for each case.

**Cluster dynamics model**

A deterministic cluster dynamics (CD) model (IRadMat) has been developed for the simulation of long-time defect evolution based on the mean-field rate theory. IRadMat has been successfully applied for the analysis of H/He retention in polycrystalline W/Be under various irradiation conditions (such as ion energy, flux, fluence, and temperature) (*44, 45, 59, 61, 66, 67*) and the mechanisms of H embrittlement in Fe (*68*). In IRadMat, the evolution of different types of defects is described by a set of reaction equations by considering the diffusion process of mobile defects with possible reactions with other defects, as given by the master equations (*44, 45, 59, 61, 66, 67*),

$$\frac{\partial C_\theta}{\partial t} = G_\theta + \sum_{\theta'}\left[\omega(\theta',\theta)C_{\theta'} - \omega(\theta,\theta')C_\theta\right] - L_\theta, \qquad (8)$$

where $C_\theta$ is the concentration of defect θ in the irradiated system at a specific time *t*. The basic types of defects θ included in Fe and W are SIAs (I), vacancies (V), and their complex clusters ($I_n$ and $V_n$, where n is the number of defects in a loop/cluster). Inherent defects such as GBs and DLs are also included in our model. In practice, only I, di-interstitials ($I_2$), and V are considered to be mobile for simplification, whereas all other defect clusters are considered to be immobile. Here, $G_\theta$ is the production rate of Frenkel pairs. $\omega(\theta',\theta)$ is the rate coefficient per unit concentration of $\theta'$-type defect/cluster transforming to $\theta$-type. $L_\theta = K_{GB}^\theta D_\theta C_\theta + K_{DL}^\theta D_\theta C_\theta$ represents the inherent absorption rate, where $K_{GB}^\theta$ and $K_{DL}^\theta$ are the sink strength of GBs and DLs, respectively. In addition, a uniform distribution of DL density is included with the corresponding sink strength (*69*), $K_{DL}^\theta = \rho Z_{DL}^\theta$, where $\rho$ is the DL density in materials and $Z_{DL}^\theta$ is a dimensionless factor representing the absorption efficiency of point defects by DLs related to the defect-dislocation elastic interaction, which is usually set as $Z_{DL}^I = 1.2$ and $Z_{DL}^V = 1.0$, respectively. The cellular model (*40*) is adopted to describe the absorption of uniform GBs to mobile defects, with the sink strength $K_{GB}^\theta$ given in Eq. (3). The reaction event list and rate coefficients are given to describe the reaction process between defects (Table S1) (*44, 61*). All parameters for defects in metals were carefully selected from the published values in experiments or DFT/MD calculations (Table S2). To further decrease the computational cost, the Fokker-Planck approximation was adopted in our model to transform these discrete master equations into continuous equations based on the Taylor expansion to the second term (*70, 71*).

Here the set of ordinary differential equations was solved using the *lsoda* subroutine package (*72*).

**Steady state rate theory model**

The RT model is used to describe the concentration of single SIAs and vacancies by considering the generation, I-V recombination and absorption of SIAs and vacancies by sinks like GBs in the steady state (*23*, *30*). Based on the RT model, the time dependence of point defect (SIA and vacancy) concentration can be given as following,

$$\partial C_{I}/\partial t = G - R_{re}C_{I}C_{V} - K_{GB}D_{I}C_{I}, \qquad (9)$$

$$\partial C_{V}/\partial t = G - R_{re}C_{I}C_{V} - K_{GB}D_{V}C_{V}, \qquad (10)$$

where $C_I$ and $C_V$ are the concentration of SIAs and vacancies, respectively. $R_{re} \approx 4\pi(D_I + D_V)/a^2$ is the recombination coefficient of Frenkel pairs, $a$ is lattice constant, and $K_{GB} = 57.6/d^2$ is the sink strength of GBs (*23*). Only GB absorption and defect self-recombination on defect concentration under different conditions are considered for simplification here. In the steady state ($\partial C_I/\partial t = \partial C_V/\partial t = 0$), the analytical solution of the vacancy concentration can be obtained, as given in Eq. (6) in the text.

# Supplementary materials for Absorption bias: An ideal descriptor for radiation tolerance of nanocrystalline BCC metals

Liuming Wei[1,2], Zhe Zhao[1], Yonggang Li[1,2], Qirong Zheng[1,2], Chuanguo Zhang[1], Jingyu Li[1], Gaofeng Zhao[3], Bo Da[4] and Zhi Zeng[1,2]

[1] Key Laboratory of Materials Physics, Institute of Solid State Physics, HFIPS, Chinese Academy of Sciences, Hefei 230031, People's Republic of China

[2] University of Science and Technology of China, Hefei, 230026, People's Republic of China

[3] Institute for Computational Materials Science, School of Physics and Electronics, Henan University, Kaifeng, 475004, People's Republic of China

[4] Research and Services Division of Materials Data and Integrated System, National Institute for Materials Science, Namiki 1-1, Tsukuba, Ibaraki 305-0044, Japan

**Supplementary Text**

**Section S1. Molecular dynamics simulation of uniform defects evolution near a grain boundary in Fe and W.**

As shown in Fig. S1, the minimum energy configurations of symmetrical twin grain boundary (GB) structure $\sum 5$ (031)-[100] ($\sum 5$ tilt) is typically selected for both body-centred cubic (BCC) Fe and W. Then, both initial self-interstitial atoms and vacancies are randomly introduced in the simulation box. The simulation system is relaxed within the NVT ensemble (constant number of atoms, volume and temperature) for 2 nanoseconds (ns) with a time step of 1 femtosecond. To reduce statistical error, 10 independent simulations of the annihilating of point defects (SIAs (I) and vacancies (V)) by GBs in Fe and W in ns are performed using a random distribution of $3\times10^3$ initial point defects for the same case.

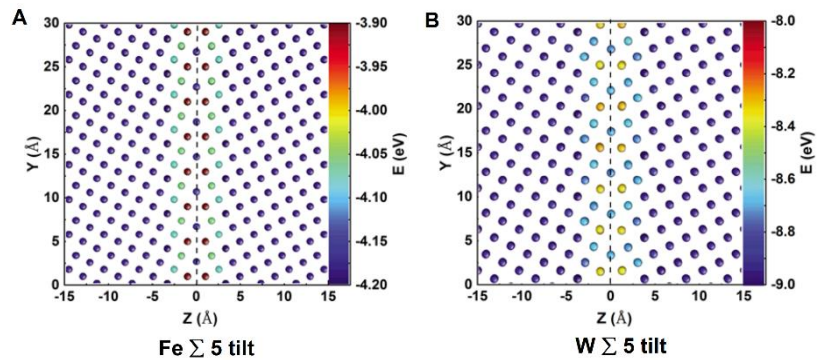

**Fig. S1. Tilt GB structures and atomic potential energies.** Stable structures of ∑5 tilt GB of (**A**) Fe and (**B**) W. The tilt GB structures are projected from the x-axis, and the z-axis is normal to the GB plane. The atomistic configurations are visualized using the AtomEye tool (*1*). Here, atoms are coloured with their potential energies, and the corresponding values are shown in the colour bar. (The simulation box contains $4\times10^5$ atoms.)

**Section S2. Experimental verification of cluster dynamics model.**

The cluster dynamics (CD) model (IRadMat) could be verified by comparing with experimental results (*2*, *3*) directly. The simulation conditions are set as pure Fe with the grain size of 30 μm and the dislocation line (DL) density of $10^{12}$ m$^{-2}$ under neutron irradiation with a flux of $3.75\times10^{-8}$ displacements per atom/s (dpa/s) at 320 K. Both experiment and CD results of the dislocation loop density increase monotonously with increasing irradiation dose, as shown in Fig. S2. Whereas the experimental results increase a little faster than that of CD when the irradiation dose is larger than ~ 0.1 dpa. This deviation can be attributed to that CD does not intrinsically take into account the spatial correlations among Frenkel pairs in cascades under high energy neutron irradiation. This would cause more I-V recombination occurs in the bulk and thus reduce the amount of dislocation loop density (*4*). In fact, the saturation behaviour of dislocation loop density is a typical theoretical trend for the deterministic stage in Fe under neutron irradiation with dose more than 0.1 dpa, which is also consistent with previous CD (*5*, *6*) and KMC simulations (*7*).

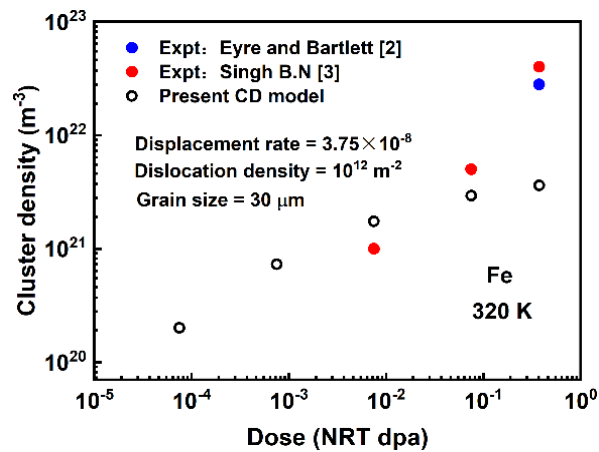

**Fig. S2. Validation of the CD model.** Comparison of calculated dose-dependent dislocation loop density in pure Fe with that of experiment values (*2*, *3*) under neutron irradiation with a flux of $3.75\times10^{-8}$ dpa/s at 320 K.

**Section S3. Vacancy concentration *v.s.* grain size.**

We take Fe and W as examples to study the effect of grain size on vacancy concentration under typical service temperatures (400 and 900 K), as shown in Fig. S3. For Fe, with increasing grain size, vacancy concentration continuously increases firstly and then declines to a constant of single crystal (SC) value at 400 K, while increases to a constant at 900 K. For W, with increasing grain size, vacancy concentration will continuously decline to constant SC value at 400 K, while increases firstly and then declines to constant SC value at 900 K. In addition, the vacancy concentration of Fe in nm scale is lower than that of SC case due to its low absorption bias. There is a typical heavy damage region (higher vacancy concentration than the SC case) of W in nm scale at low temperatures due to the effect of high absorption bias becomes extremely serious. Therefore, nanocrystalline (NC) Fe has good irradiation resistance at 400 K and 900 K, while NC W has good irradiation resistance only at 900 K, but weaker than coarse-grained (CG) W at 400 K.

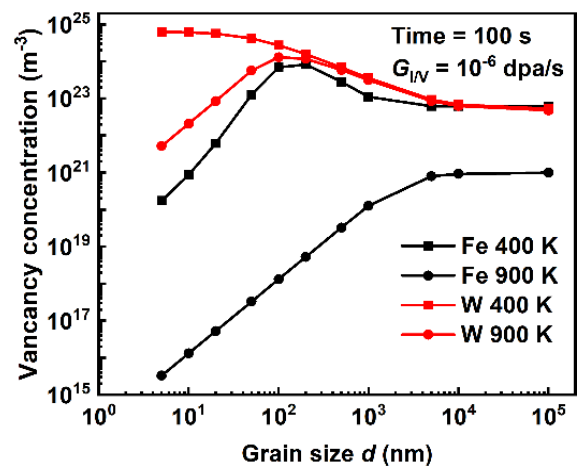

**Fig. S3. Vacancy concentration *v.s.* grain size.** Grain size dependence of vacancy concentration in Fe and W under uniform irradiation of 100 s with the production rate of Frenkel pairs of $10^{-6}$ dpa/s at 400 and 900 K.

**Section S4. Defect annealing by cluster dynamics.**

The same trends of vacancy concentration under irradiation and after a long-time (1000 s) annealing without continuing irradiation at 300 K will continuously decline to a constant SC value with increasing grain size. The radiation tolerance of NC W barely changes with increasing grain size after annealing at 300 K, which is mainly due to the barely immobile vacancies with high migration energy of 1.66 eV. However, when the temperature increases to 900 K, vacancy diffusion is activated and thus the advantage of NC W gradually becomes prominent (Fig. S4). With increasing grain size, vacancy concentration from a very low value increases to a constant SC value at 900 K after annealing. W with grain size $d_1$ (a constant value after reaching an inflection point) changing from about 20 nm under irradiation to several hundred nm after 1000 s annealing exhibit better irradiation resistance than CG W. However, under realistic irradiation conditions, nuclear materials will face substantial and sustainable ion and neutron bombardment. Thus, the concentration of defects is always at a very high level. In this case, the dynamic evolution of defects remains far from the steady state. The level of diffusion bias and absorption bias should still dominate the radiation tolerance of base materials and materials with different grain sizes under typical service conditions in fusion devices.

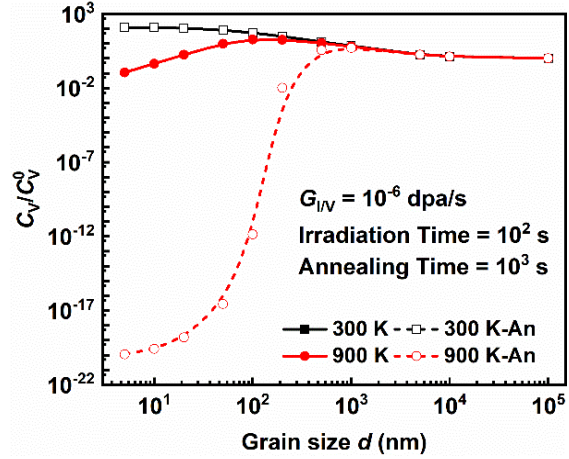

**Fig. S4. Vacancy concentrations after annealing.** Grain size dependence of normalized vacancy concentration in W under uniform irradiation of 100 s with the production rate of Frenkel pairs of $10^{−6}$ dpa/s annealing 1000 s at 300 and 900 K.

**Section S5. Equilibrium time of mobile defects in Fe and W under different conditions**

When the system is in the steady state, both NC Fe and W show better radiation tolerance compared to their CG counterparts. We will see that under typical service conditions in fusion devices, the steady state cannot always reach of CG Fe and W at 300 and 900 K, respectively. According to the relation $d \sim \sqrt{4D_\theta t}$ that the diffusion distance ($d$) from grain interior to GBs for a vacancy with the diffusion coefficient $D_\theta$ after time $t$, the equilibrium time of vacancy evolution in Fe and W without continuing point defect production is estimated, as shown in Fig. S5. The equilibrium time increases with increasing grain size and decreases with increasing temperature. Vacancies in NC Fe and W with the grain size of 100 nm need about $10^2$ and $10^{20}$ s to reach the equilibrium state at 300 K, respectively. However, for W selected in ITER or commercial W with the grain size about 10 μm at high temperature of 900 K, it still needs a long time (~ $10^7$ s) to reach its equilibrium state.

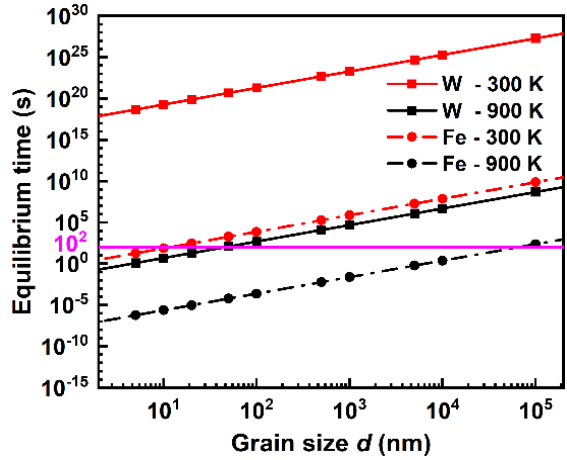

**Fig. S5**. **Equilibrium time of vacancy diffusion.** Relationship of equilibrium time with grain size for Fe and W at two typical temperatures of 300 K and 900 K.

**Section S6. Critical grain size of anti-irradiation in the steady state.**

As shown in Fig. S6, there are two asymptotes for the change of $C_V$ with $d$ in the limits of small grain size and SC that correspond to the contributions of fully GB absorption and I-V recombination, respectively. The intersection of these two asymptotes is defined as the critical grain size ($d_c$) (as shown in the insert Fig. S6). The $d_c$ of metals with high migration energies (Ta, Mo and W) are much smaller than those of metals with low migration energies (Fe, V, Nb and Cr). It is mainly because of the high vacancy migration energies and low absorption rates of vacancies by GBs in Ta, Mo and W. When the grain size becomes larger, vacancies with a smaller diffusivity are hard to be absorbed by GBs, so the corresponding vacancy concentration tends to that in SC. Therefore, the larger vacancy diffusion coefficient, the larger $d_c$ of material is. $d_c$ can thus be used to represent the anti-irradiation ability of materials under different service conditions, that is, larger $d_c$ means higher radiation resistance. Besides BCC metals, hexagonal close-packed (HCP) and face-centred cubic (FCC) metals also follow the relationship of Eq. (7) under irradiation with the production rate of Frenkel pairs of $10^{-6}$ dpa/s at 600 K. In general, HCP and FCC metals have high radiation resistance. For BCC metals, the range of $d_c$ is from a few nm (W) to several hundred nm (Fe). As can be seen in Eq. (7), $d_c$ is mainly determined by vacancy diffusivity $D_V$ (where $D_V$ is related to vacancy migration energy $E_V^m$ and temperature $T$) and the production rate of Frenkel pairs $G$. Therefore, by selecting base metals with appropriate $D_V$ (or $E_V^m$), new materials with high radiation resistance could be designed by tuning $d$ to endure different service environments (like $G$ and $T$). Under the steady state, a high radiation resistant material should have low $E_V^m$ and small $d$. Meanwhile, the lower $G$ and higher $T$, the better anti-irradiation ability of the material.

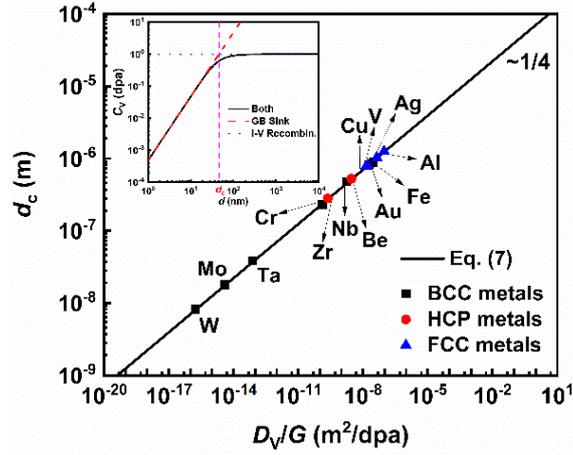

**Fig. S6. Anti-irradiation evaluation in the steady state.** The relationship between $d_c$ and $D_V/G$, for 13 BCC, HCP and FCC metals under uniform irradiation with the production rate of Frenkel pairs of $10^{-6}$ dpa/s at 600 K. The insert figure shows the grain size-dependent vacancy concentration $C_V$ (black solid line) normalized to the SC $C_V^0$ (blue dot line), and the vacancy concentration for materials with different grain sizes (red dash line).

**Section S7. Defect behaviour by steady state rate theory.**

We take Fe and W as examples to study the effect of grain size on vacancy concentration under different temperatures in the steady state, as shown in Fig. S7. The relative vacancy concentration ($C_V/C_V^0$) of Fe and W decreases with decreasing grain size and increasing temperature, indicating that the anti-irradiation ability of Fe and W appears for NCs and at high temperatures in the steady state. The anti-irradiation region (up the white line of $d_c$) in Fe is larger than that in W, implying that the radiation tolerance of Fe is much higher than that of W because of its higher vacancy diffusivity at the same service conditions. The anti-irradiation region appears at grain sizes in the micrometre range in Fe but at grain sizes of several hundreds of nanometres in W, implying that the radiation tolerance in Fe is much higher than that in W under the same service conditions.

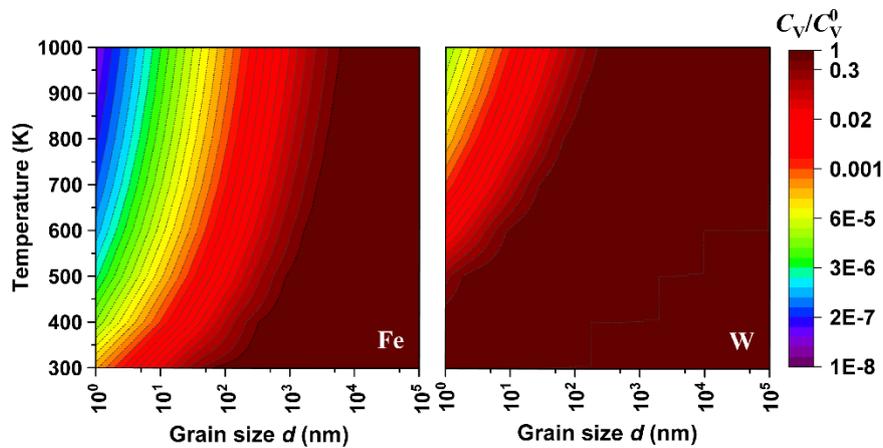

**Fig. S7.** Anti-irradiation window prediction in the steady state. Grain size and temperature dependence of relative vacancy concentration in Fe and W in the steady state. The colour scale

indicates the relative vacancy concentration ($C_V/C_V^0$) to that of SC ($C_V^0$).

**Section S8. Sink strength *v.s.* grain size.**

Taking W as examples, we study the effect of grain size on the sum of the sink strengths of all the intragranular sinks but the GB ($k_{sc}$) and sink strength of GB ($K_{GB}$) at 400 K, as shown in Fig. S8. Since the density of DLs ($\rho$) is constant, the difference between $k_{sc}^I$ and $k_{sc}^V$ mainly depends on the sink strengths of single defects and their clusters. With increasing $d$, $k_{sc}^I$ is essentially unchanged since the sink strength of single defects and their clusters are low compared to DLs, while the $k_{sc}^V$ will increase firstly and then tend to SC with increasing grain size since the sink strength of single defects and their clusters are high compared to DLs. The larger vacancy migration energy, the smaller vacancy diffusion coefficient is, and thus more rapidly change of $k_{sc}^V$ is. With increasing $d$, the $K_{GB}^I$ will continuously decrease, $K_{GB}^V$ decreases firstly at the same tend as $K_{GB}^I$ then slight increases after $d$ reaching the inflection point $d_1$, and then tends to decrease in the same tend as $K_{GB}^I$ after $d$ reaches the other inflection point $d_2$ ($d_1$ and $d_2$ are defined in the discussion of text). When $d < d_1$, SIAs and vacancies can fully interact with GBs to reach their respective stable states of maximum absorption. When $d_1 < d < d_2$, on the one hand, with increasing $d$, the effect of $d$ itself on the change of GB absorption abilities to SIAs and vacancies are reduced more and less. On the other hand, $k_{sc}$ contains the sink strengths of DLs ($K_{DL}$) as well as single defects and their clusters that are proportional to sink densities and rate coefficients. Therefore, the synergistic effect of $d$ and $k_{sc}$ causes the dramatical decrease in $K_{GB}^I$ but slight increase in $K_{GB}^V$, thus the reduction of absorption bias ($B_A$). When $d > d_1$, $k_{sc}^I$ and $k_{sc}^V$ does not change with grain size anymore since the sink strength of single defects and their clusters gradually weaken under the influence of GBs and eventually tends to SC state.

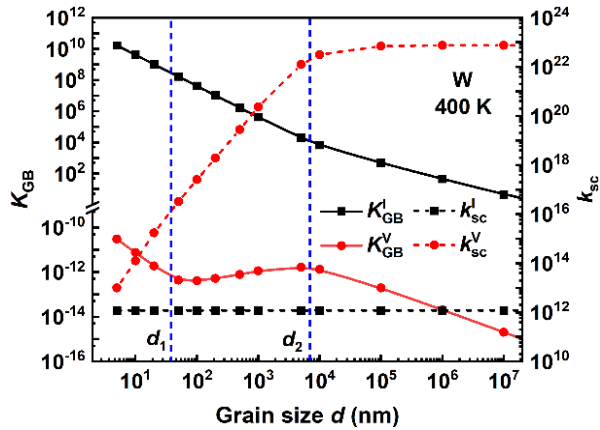

**Fig. S8. Sink strength.** Sink strength of $K_{GB}$ and $k_{sc}$ as a function of grain size of W under uniform irradiation of 100 s with the production rate of Frenkel pairs of $10^{-6}$ dpa/s at 400 K.

**Section S9. Reaction event list and rate coefficients in cluster dynamics.**

The reaction event list and their corresponding rate coefficients in BCC metals (Table S1) are set the same values as those in our previous papers (*8*, *9*), and their reliability has been widely verified. Typical data are provided in Table S2.

**Table S1. Reaction event list and corresponding rate coefficients.**

| Reaction events | Rate coefficients |
|---|---|
| $I + V \rightleftharpoons 0$ | $k^+_{I+V}$, $G_{I/V}$ |
| $I + I_n \rightleftharpoons I_{n+1}$ | $\alpha^+_n$, $\alpha^-_{n+1}$ |
| $I + V_n \rightarrow V_{n-1}$ | $k^+_{V_n+I}$ |
| $I_2 + I_n \rightleftharpoons I_{n+2}$ | $\beta^+_n$, $\beta^-_{n+2}$ |
| $I_2 + V_n \rightarrow V_{n-2}$ | $k^+_{V_n+I_2}$ |
| $V + I_n \rightleftharpoons I_{n-1}$ | $k^+_{I_n+V}$, $k^-_{I_{n-1}-V}$ |
| $V + V_n \rightleftharpoons V_{n+1}$ | $\gamma^+_n$, $\gamma^-_{n+1}$ |
| $\theta + GB \rightarrow GB\theta$ | $k^+_{GB+\theta}$, $\theta = I, I_2, V$ |
| $\theta + DL \rightarrow DL\theta$ | $k^+_{DL+\theta}$, $\theta = I, I_2, V$ |

**Table S2. Parameters used for various BCC metals under neutron irradiation.**

| Parameters | Symbol and Value | V (*10–12*) | Cr (*10, 13,14*) | Fe (*15,16*) | Nb (*10,17*) | Mo (*10, 18,19*) | Ta (*10, 12,13*) | W (*20*) |
|---|---|---|---|---|---|---|---|---|
| Lattice constant | (Å) | 3.04 | 2.885 | 2.87 | 3.32 | 3.17 | 3.31 | 2.885 |
| Recombination | (Å) | 2.7 | 2.54 | 6.50 | 2.96 | 2.8 | 2.96 | 2.54 |
| SIA pre-exponential factor | (m²s⁻¹) | $10^{-7}$ | $1.38 \times 10^{-7}$ | $6.20 \times 10^{-8}$ | $2.00 \times 10^{-6}$ | $1.6 \times 10^{-5}$ | $1.83 \times 10^{-7}$ | $10^{-8}$ |
| V pre-exponential factor | (m$^2$s$^{-1}$) | 10$^{-7}$ | 8.29×10$^{-7}$ | 6.2×10$^{-8}$ | 1.15×10$^{-5}$ | 3.0×10$^{-6}$ | 1.09×10$^{-6}$ | 10$^{-8}$ |
| Migration energy of V | (eV) | 0.70 | 0.95 | 0.67 | 0.83 | 1.5 | 1.35 | 1.66 |
| Migration energy of SIA | (eV) | 0.03 | 0.0953 | 0.34 | 0.049 | 0.085 | 0.022 | 0.013 |
| Formation energy of V | (eV) | 2.06 | 2.64 | 2.07 | 2.75 | 2.96 | 2.83 | 9.466 |
| Formation energy of SIA | (eV) | 2.97 | 5.66 | 3.77 | 4.349 | 7.34 | 5.08 | 3.80 |
| Binding energy of V$_2$ | (eV) | 0.31 | 0.2225 | 0.30 | 0.36 | 0.268 | 0.31 | 0.656 |
| Binding energy of I$_2$ | (eV) | 0.71 | 1.197 | 0.80 | 0.994 | 0.349 | 1.23 | 2.12 |